\documentclass[aps,prl,twocolumn, superscriptaddress,groupedaddress]{revtex4}

\usepackage{graphicx}
\usepackage{amsmath}
\usepackage{amssymb}
\usepackage[amssymb]{SIunits}
\usepackage{color}
\usepackage[usenames,dvipsnames]{xcolor}
\usepackage{ulem}
\usepackage{cancel}
\usepackage{natbib}
\usepackage[T1]{fontenc}
\usepackage{bm}

\newcommand{\dr}[1]{\ifmmode\text{\textcolor{blue}{\sout{\ensuremath{#1}}}}\else\textcolor{blue}{\sout{#1}}\fi}

\begin{document}

\title{Supergravitational turbulent thermal convection}

\author{Hechuan Jiang}
\thanks{These authors contributed equally to this work}
\affiliation{Center for Combustion Energy, Key Laboratory for Thermal Science and Power Engineering of MoE, and Department of Energy and Power Engineering, Tsinghua University, 100084 Beijing, China.}
\affiliation{Department of Engineering Mechanics, School of Aerospace Engineering, Tsinghua University, Beijing 100084, China}

\author{Xiaojue Zhu}
\thanks{These authors contributed equally to this work}
\affiliation{Center of Mathematical Sciences and Applications, and School of Engineering and Applied Sciences, Harvard University, Cambridge, MA 02138, USA}

\author{Dongpu Wang}
\affiliation{Center for Combustion Energy, Key Laboratory for Thermal Science and Power Engineering of MoE, and Department of Thermal Engineering, Tsinghua University, 100084 Beijing, China.}
\affiliation{Department of Engineering Mechanics, School of Aerospace Engineering, Tsinghua University, Beijing 100084, China}

\author{Sander G. Huisman}
\affiliation{Physics of Fluids Group, Max Planck UT Center for Complex Fluid Dynamics, MESA+ Institute and J.M. Burgers Centre for Fluid Dynamics, University of Twente, P.O. Box 217, 7500 AE Enschede, The Netherlands}

\author{Chao Sun}
\thanks{chaosun@tsinghua.edu.cn}
\affiliation{Center for Combustion Energy, Key Laboratory for Thermal Science and Power Engineering of MoE, and Department of Thermal Engineering, Tsinghua University, 100084 Beijing, China.}
\affiliation{Department of Engineering Mechanics, School of Aerospace Engineering, Tsinghua University, Beijing 100084, China}

\date{\today}

\begin{abstract} 
{\bf
Teaser: A novel system using supergravity opens a new avenue on the exploration of high Rayleigh number thermal turbulence.}
\bigskip

{\bf High-Rayleigh number convective turbulence is ubiquitous in many natural phenomena and in industries, such as atmospheric circulations, oceanic flows, flows in the fluid core of planets, and energy generations. In this work, we present a novel approach to boost the Rayleigh number in thermal convection by exploiting centrifugal acceleration and rapidly rotating a cylindrical annulus to reach an effective gravity of 60 times Earth's gravity. We show that in the regime where the Coriolis effect is strong, the scaling exponent of Nusselt number versus Rayleigh number exceeds one-third once the Rayleigh number is large enough. The convective rolls revolve in prograde direction, signifying the emergence of zonal flow. The present findings open a new avenue on the exploration of high-Rayleigh number turbulent thermal convection and will improve the understanding of the flow dynamics and heat transfer processes in geophysical and astrophysical flows and other strongly rotating systems.}
\end{abstract}

\maketitle
\section{Introduction}

Thermally driven turbulent flows widely occur in geophysical flows and industrial processes. Examples are thermal convection in the atmospheric and mantle convection \cite{mck1974,Wyngaard1992}, in the ocean \cite{cheng2019fast}, and in many industrial processes \cite{Bejan}. Rayleigh--B\'enard convection (RBC), a fluid layer heated from below and cooled from above, is an ideal model for the study of thermally driven turbulent flows \cite{ahl09,loh10,chi12}. 
The main challenge for thermal turbulence studies is to explore the flow dynamics and heat transfer of the system in a wide range of the control parameters. The main attention has been on how the heat transfer depends on the Rayleigh number,  which is the dimensionless temperature difference and measures the intensity of the thermal driving of the system. The Rayleigh number is defined as $\text{Ra}=\frac{\beta g \Delta L^3 }{\nu \kappa}$, where $g$ is the gravitational acceleration, $\beta$ is the isobaric thermal expansion coefficient, $\nu$ the kinematic viscosity, $\kappa$ is the thermal diffusivity, $\Delta$ is the temperature difference between the hot plate and the cold plate, and $L$ is the thickness of the fluid layer between the aforementioned plates.  

To achieve high Rayleigh numbers, numerous strategies have been proposed in the past decades. Looking at the definition of $\text{Ra}$, one can see that large $\text{Ra}$ can be reached in several ways. 
One approach is to use a working fluid for which the parameter $\frac{\beta}{{\nu}\kappa}$ has a large value. This approach has been widely used in the community, such as using helium gas at cryogenic temperatures \cite{cas89,cha97,nie00,Urban2010} or mercury \cite{san89}. Recently, it was found that pure gas, particularly gasses with high molecular weight and under high pressure (up to 19-bar), is also an effective way to push $\text{Ra}$ to high values at a constant $\text{Pr}$ number \cite{fun09,ahl09b}. Another commonly used approach in the field of turbulent thermal convection is to increase the system thickness $L$ \cite{nie00,fun05,nik05,sun05e,pui07,fun09,ahl09b,Urban2010}. 

The fact that hitherto not much attention was on the changing  the gravitational acceleration provides the motivation of the present work. In this study, we propose a novel system, annular centrifugal RRC (ACRBC), which is a cylindrical annulus with cooled inner and heated outer walls under a solid body rotation. 
In this system, the buoyancy force can be efficiently enhanced by replacing the gravitational acceleration ($g$) by the centrifugal acceleration, and  consequently, $\text{Ra}$ can be increased by increasing the rotation rate for a given fluid and temperature difference. 
Another advantage of this system is that the thermal convection in rapidly rotating cylindrical annulus has been recognized as a good model for the study of the flows in planetary cores and stellar interiors \cite{hid58,bus74,hid75,bus76,aue95,kan19} and also in engineering applications under strong rotation, \textit{i.e.}~flow in gas turbines \cite{mar05,mic15,pit17}. The flow dynamics in this system can give insights into flows in geophysical context and flows in industrial processes. In this work, we aim to study the heat transfer and the turbulent structures in ACRBC.

\section{Governing equations, simulations, and experiments}
\subsection{Governing equations and numerical simulations}

\begin{figure*}
\centering
\includegraphics[width=1\linewidth]{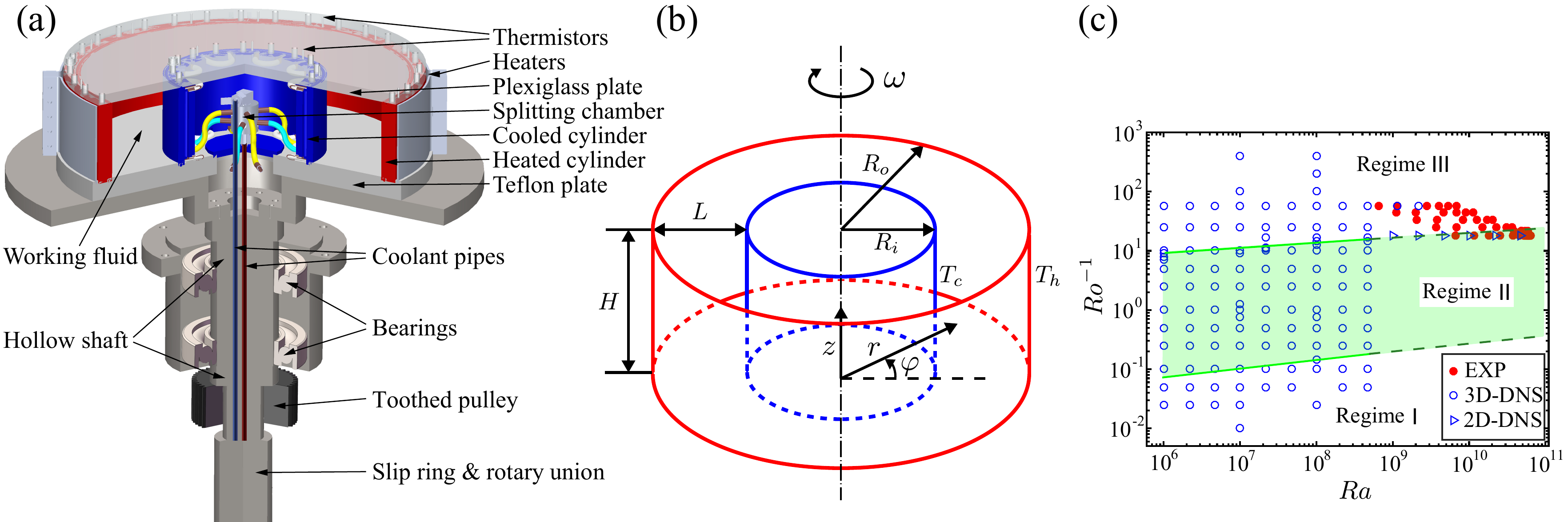}
\caption{{\bf System configuration and parameter space.} (a) Three-dimensional render of the experimental set-up. A fluid is confined in a rotating cylindrical annulus having its inner(blue) and outer(red) cylindrical surfaces made out of copper, which is known for its excellent thermal conductivity. The bottom plate is made of teflon for thermal insulation, and top plate is made of plexiglass allowing flow visualization. The coolant pipes and electric cables go through the hollow stainless steel shaft and connect to a slip ring and a rotary union. The toothed pulley is driven by a servo motor. (b) Schematical diagram of the set-up, which defines the geometric parameters. (c) Explored parameter space of $\text{Ra}$ and $\text{Ro}^{-1}$ for our study of supergravitational thermal convection. The red discs, the blue circles and the blue triangles correspond to the parameters used in the experiments (EXP), in the three-dimensional (3D) numerical simulations and in the two-dimensional (2D) numerical simulations, respectively. The parameter space is divided into three regimes according to the influence of Coriolis force. In addition, for comparison we have also performed the simulations for $\text{Ro}^{-1}=10^{-5}$ (not shown here), for which the Coriolis force is negligible. For more details about the experimental apparatus, we refer to the Supplementary Materials and movie S1.}
\label{fig:setup}
\end{figure*}

Using the parameter definitions as shown in Fig.~\ref{fig:setup}(b), the governing dimensionless Boussinesq equations in a rotating reference frame \cite{lop13} can be expressed as
\begin{small}
\begin{align}
 \nabla \cdot {\bf u} &= 0 \\
\frac{\partial \theta}{\partial t} +{\bf u}\cdot \nabla \theta &= \frac{1}{\sqrt{RaPr}}\nabla^2 \theta \\
\frac{\partial \bf{u}}{\partial t} +{\bf u}\cdot \nabla {\bf u} &= -\nabla p+\text{Ro}^{-1} \hat{\bm{\omega}} \times {\bf u} \nonumber \\ 
&\phantom{=} +\sqrt{\frac{Pr}{Ra}}\nabla^2 {\bf u} -\theta \frac{2(1-\eta)}{(1+\eta)}\bf r
\end{align}
\end{small}
\noindent where $\bm{\hat{\omega}}$ is the unit vector pointing in the direction of the angular velocity, $\bf u$ is the velocity vector normalized by the free fall velocity $\sqrt{\omega ^2\frac{R_o+R_i}{2}\beta \Delta L}$, $t$ is the dimensionless time normalized by $\sqrt{L/(\omega ^2\frac{R_o+R_i}{2}\beta\Delta)}$, and $\theta$ is the temperature normalized by $\Delta = T_h - T_c$. Here, $\omega$, $R_i$, $R_o$, and $\Delta$ are the rotational speed, the thickness of the fluid layer between the two cylinders, the outer radius of the inner cylinder, the inner radius of the outer cylinder, and the temperature difference between the hot ($T_h$) and cold ($T_c$) cylinders, respectively.

From the above governing equations, the relevant control parameters in ACRBC are the Rayleigh number (characterizing the thermal driving strength) 
$Ra=\frac{1}{2}\omega^2(R_o+R_i)\beta \Delta  L^3/(\nu \kappa)$,
the inverse Rossby number (measuring Coriolis effects)
$Ro^{-1}=2(\beta \Delta (R_o+R_i)/(2L))^{-1/2}$,
the Prandtl number (fluid property) $\text{Pr}=\nu/\kappa$, and the radius and aspect ratio (geometric properties) $\eta=R_i/R_o$ and $\Gamma=H/L$. The key response parameter of the system is the Nusselt number (measuring the ratio of the total heat transport over the conductive one) $\text{Nu}=-Q\ln(\eta)/(\alpha \Delta 2 \pi H)$. Here, $Q$ is the measured heat input through the outer cylinder into the system per unit of time; $\alpha=\kappa\rho c_p$ is the thermal conductivity of the working fluid with $\rho$ and $c_p$ being the density and the specific heat capacity of the fluid, respectively; and $H$ is the height of the gap between two cylinders. It should be noted that the definition of $\text{Nu}$ in ACRBC is slightly different from that in classical RBC because of the cylindrical geometry with a heat flux in the radial direction, and the detailed derivations of the $\text{Nu}$ for ACRBC are documented in the Supplementary Materials.

Direct numerical simulations (DNS) are performed using an energy conserving second-order finite-difference code \cite{ver96,poe15cf,zhu18afid}. The code has been extensively validated and used in previous studies in both the Cartesian and cylindrical coordinates for convective systems \cite{poe15cf,zhu18afid,zhu18prl,zhu18np}. In all numerical simulations and experiments, the radius ratio is fixed at $\eta=0.5$. No-slip boundary conditions were used for the velocity and constant temperature boundary conditions for the inner and outer cylinders. Periodic boundary conditions were used for the top and bottom surfaces. For three-dimensional simulations, the aspect ratio was chosen the same as in the experiments $\Gamma=H/L=1$ (the only two exceptions are for the cases at $\text{Ra}=1.16\times10^9$ and $\text{Ra}=2.2\times10^9$, where the height of the domain is $L/4$ and $L/8$, respectively, as the flows here are quasi two-dimensional). For the cases of which the flows are quasi two-dimensional at large  $\text{Ra}$, we use two-dimensional simulations. Pr was fixed at 4.3 for all the simulations. The adequate resolution was ensured for all cases, for example, at $\text{Ra}=4.7 \times 10^8$, $4608\times384\times384$ grid points were used and at  $\text{Ra}=4.7 \times 10^{10}$ with grid points $18432\times1536$. To obtain sufficient statistics, each simulation was run at least 200 free fall time units. As reported in \citep{kun10,kun11}, the boundary layers in rotation system are expected to be thiner compared with classical RBC. Hence, the verification of boundary layer grid resolution is necessary. A posteriori check of grid resolution shows that at $\text{Ra}=4.7\times10^{10}$, there are 48 grid points inside thermal boundary layers and 64 grid points inside viscous boundary layers, which guarantees to resolve the boundary layer adequately. Besides, we have conducted a set of grid independence studies and checked the spatial resolution in bulk region to resolve all relevant scales (Kolmogorov scale and Batchelor scale). The parameter space that was explored can be found in Fig. \ref{fig:setup}(c). More details about the DNS are documented in the Supplementary Materials.

\bigskip

\subsection{Experiments}
Experiments are performed in a cylindrical annulus with solid-body rotation as sketched in Fig.~\ref{fig:setup} (a). The two concentric cylinders are machined from a solid piece of copper to control the symmetry of the system, and their surfaces are electroplated with a thin layer of nickel to prevent oxidation. The inner cylinder has an outer radius of $R_i=\unit{120}{\milli\meter}$, and the outer cylinder has an inner radius of $R_o=\unit{240}{\milli\meter}$, resulting in a gap of L=$\unit{120}{\milli\meter}$, and a radius ratio of $\eta = R_i/R_o = 0.5$. The gap with a height $H=\unit{120}{\milli\meter}$ is sandwiched by a top plexiglass plate and a bottom teflon plate, resulting in an aspect ratio of $\Gamma = H/L = 1$. Plexiglass with an excellent transparency is used as the material of the top lid, allowing us to visualize the flow field, whereas teflon with an excellent corrosion resistance and a good strength is machined as the bottom base. These end plates and cylinders are fixed together and leveled on a rotating aluminum frame with a rotation rate up to \unit{546}{rpm}. Four silicone rubber film heaters are attached to the outside of the outer cylinder, and are supplied by a DC power supply (Ametek, \texttt{XG 300-5}) with 0.005\% long-term stability. We use water as our working fluid, which has $\text{Pr} \approx 4.3$.

Inside the cold inner cylinder, 16 channels are machined for the coolant fluid to pass through, and the inner cylinder is regulated at constant temperature using a temperature-controlled circulating bath (PolyScience, \texttt{AP45R-20-A12E}). The coolant pipes and electric cables go through the hollow stainless steel shaft and connect with a slip ring and a rotary union (Moflon, \texttt{MEPH200}), which has 2 channels for liquids, 6 channels for powers (\unit{2220}{\watt}), and 48 channels for electrical signals. The shaft is driven by a toothed belt and the pulley has a gear ratio of $60:32\approx1.88:1$. The whole system is powered by a servomotor (Yaskawa, \texttt{SGM7G-1AA}), which has a rated power of \unit{11}{\kilo\watt}.

For high-precision temperature and heat flux measurements, it is essential to minimize the heat leakage from the experimental apparatus to the surroundings. Various thermal shields that are regulated at appropriate temperatures are installed in the system to prevent heat losses.  Furthermore, the whole system is placed in a big toughened glass box where the temperature is controlled by a Proportional-Integral-Derivative (PID) controller to match the mean temperature of the bulk fluid. For more detailed descriptions on the experimental set-up, we refer to Supplementary Materials. The measurement techniques are documented in Methods.

\begin{figure*} [ht]
\centering
\includegraphics[width=\linewidth]{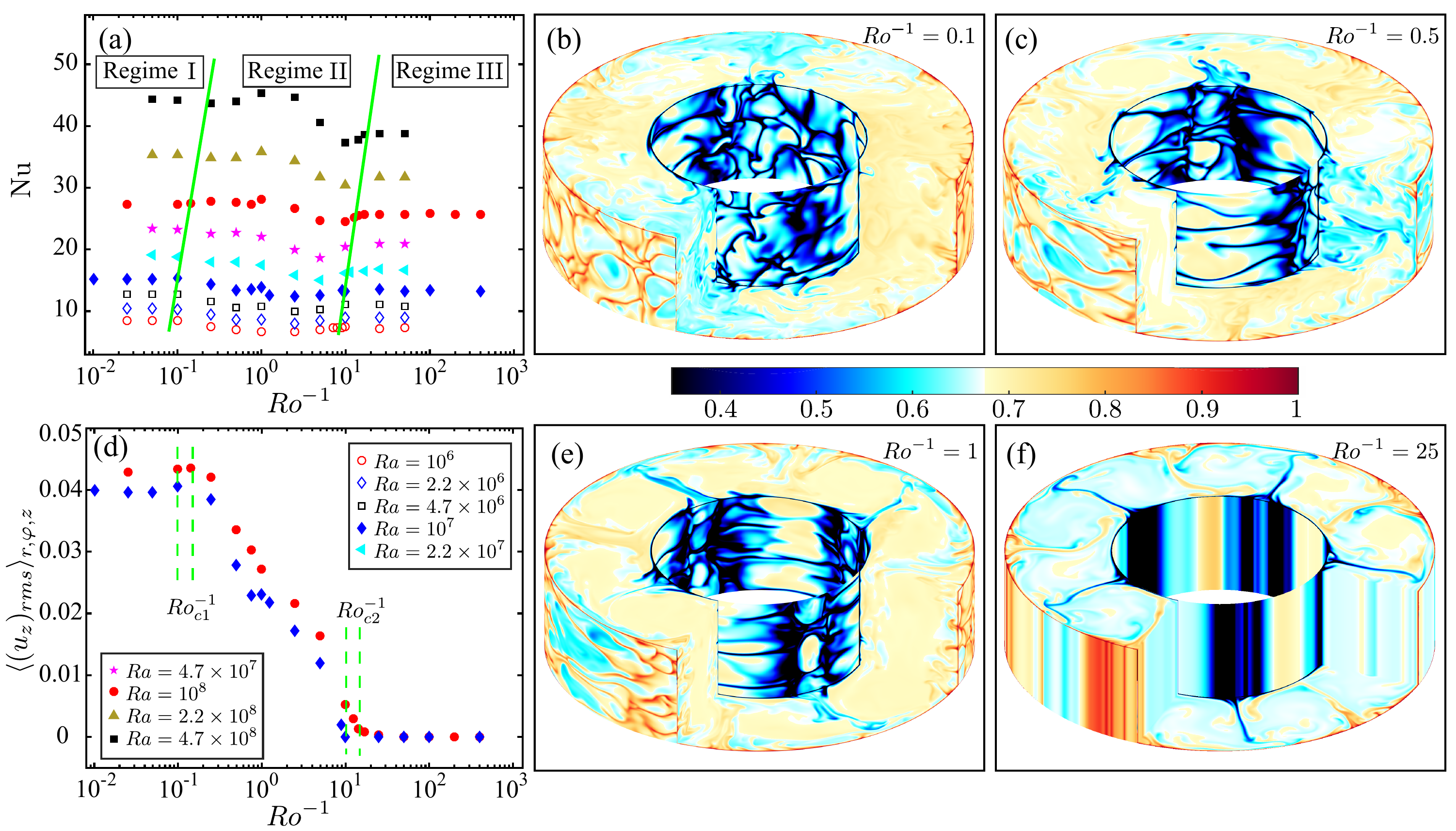}
\caption{{\bf Effects of the Coriolis force on the heat transfer and flow structures.} (a) Nusselt number ($\text{Nu}$) as a function of $\text{Ro}^{-1}$ for $\text{Ra}=10^6$, $2.2\times10^6$, $4.7\times10^6$, $10^7$, $2.2\times10^7$, $4.7\times10^7$, $10^8$, $2.2\times10^8$ and $4.7\times10^8$ for $\text{Pr} = 4.3$. (d) Root mean square axial velocity fluctuation $\langle (u_z)_{rms}\rangle _{r,\varphi, z}$ versus $Ro^{-1}$ for $\text{Ra}=10^7$ and $\text{Ra}=10^8$. (b,c,e,f) Instantaneous temperature fields from DNS at $\text{Ra}=10^8$ for $\text{Ro}^{-1}=0.1$, $0.5$, $1$, and $25$ ($\text{Pr} = 4.3$). The inner and outer surfaces locate 0.02 $L$ away from the cold and hot cylinder correspondingly.}
\label{fig:Ro}
\end{figure*}

\section {Results}

\subsection{Effects of the Coriolis force on the heat transport and flow structures}
A series of numerical simulations are performed to explore the influence of the Coriolis force on the heat transfer and flow structures. In numerical experiments, $\text{Ro}^{-1}$ varies from $10^{-2}$ to $4\times 10^2$ with several fixed Rayleigh numbers $\text{Ra}=10^{6}$, $2.2\times10^6$, $4.7\times10^6$, $10^7$, $2.2\times10^7$, $4.7\times10^7$, $10^8$, $2.2\times10^8$, and $4.7\times10^8$. As shown in Fig.~\ref{fig:Ro} (a), for these $\text{Ra}$ studied, the influence of Coriolis force can be divided into three regimes. In regime \uppercase\expandafter{\romannumeral1} ($\text{Ro}^{-1}<\text{Ro}^{-1}_{c1}$), the Coriolis force is small and negligible as compared to the buoyancy so that the $\text{Nu}$ does not depend on $\text{Ro}^{-1}$. In regime  \uppercase\expandafter{\romannumeral3} ($\text{Ro}^{-1}>\text{Ro}^{-1}_{c2}$), the rotation is so strong that the flow is nearly constrained by the effect of the Taylor-Proudman theorem~\cite{pro16,tay22}, resulting in a quasi-two-dimensional flow state and thus a reduction of $\text{Nu}$ compared with regime \uppercase\expandafter{\romannumeral1}. For regime \uppercase\expandafter{\romannumeral2} ($\text{Ro}^{-1}_{c1}<\text{Ro}^{-1}<\text{Ro}^{-1}_{c2}$), the flow is governed by the combination of $\text{Ra}$ and $\text{Ro}^{-1}$ with rich flow states. In this regime, the general trend is that Nu decreases with $\text{Ro}^{-1}$. The instantaneous temperature fields for several $\text{Ro}^{-1}$ at $\text{Ra}=10^8$ (see Fig.~\ref{fig:Ro}(b,c,e,f)) from DNS support the explanation of this $\text{Ro}$ dependence on the heat transport. As illustrated in Fig.~\ref{fig:Ro}(b), the flow is three-dimensional at $\text{Ro}^{-1}=0.1$ because the Coriolis force is too small to influence the flow effectively. However, with $\text{Ro}^{-1}$ increasing, the enhanced Coriolis force tends to suppress vertical variation of the convection flow, and the flow gradually becomes two-dimensional, which is a manifestation of the Taylor-Proudman theorem. The two-dimensionalization of the flow field should be responsible for the reduction of heat transport at $\text{Ro}^{-1} > \text{Ro}^{-1}_{c1}$, which was also reported in \cite{poe13}. As is evident in Fig.~\ref{fig:Ro} (f), the flow is nearly two-dimensional at $\text{Ro}^{-1} \gtrsim \text{Ro}^{-1}_{c2}$. So, further increased $\text{Ro}^{-1}$ saturates the influence of the Coriolis force, which explains the nearly constant $\text{Nu}$ when $\text{Ro}^{-1} \gtrsim \text{Ro}^{-1}_{c2}$. Moreover, we use root mean square axial velocity fluctuation $\langle (u_z)_{rms}\rangle _{r,\varphi, z}$ to measure the influence of the Taylor-Proudman theorem, as shown in Fig.~\ref{fig:Ro}(d). It shows that $\langle (u_z)_{rms}\rangle _{r,\varphi, z}$ has nearly a constant value when $\text{Ro}^{-1}<\text{Ro}^{-1}_{c1}$, then gradually decreases with $\text{Ro}^{-1}$, and finally approaches to around zero after  $\text{Ro}^{-1}>\text{Ro}^{-1}_{c2}$. The overall trend of $\langle (u_z)_{rms}\rangle _{r,\varphi, z}$ versus $\text{Ro}^{-1}$ is consistent with the dependence of $\text{Nu}$ on $\text{Ro}^{-1}$ in general.

We have noticed that the critical $\text{Ro}^{-1}_{c1}$ and $\text{Ro}^{-1}_{c2}$ depend on the $\text{Ra}$. As illustrated in Fig.~\ref{fig:Ro} (d), for $\text{Ra}=10^7$, we have $\text{Ro}^{-1}_{c1}\approx0.1$ and $\text{Ro}^{-1}_{c2}\approx10$, while for $\text{Ra}=10^8$, we have $\text{Ro}^{-1}_{c1}\approx0.15$ and $\text{Ro}^{-1}_{c2}\approx15$.  According to the numerical data explored, we can roughly determine the boundaries of the three regimes, which is plotted in Fig.~\ref{fig:Ro} (a) with green lines. In addition, we have also included the boundaries in Fig.~\ref{fig:setup} (c), and extrapolated the boundaries to the parameter space of experiments. (More detailed discussions of the flow structures, $\langle (u_z)_{rms}\rangle _{r,\varphi, z}$, and $\text{Ro}^{-1}_{c}$ for these $\text{Ra}$ cases are available in Supplementary Materials.)

\begin{figure*} [ht]
\centering
\includegraphics[width=\linewidth]{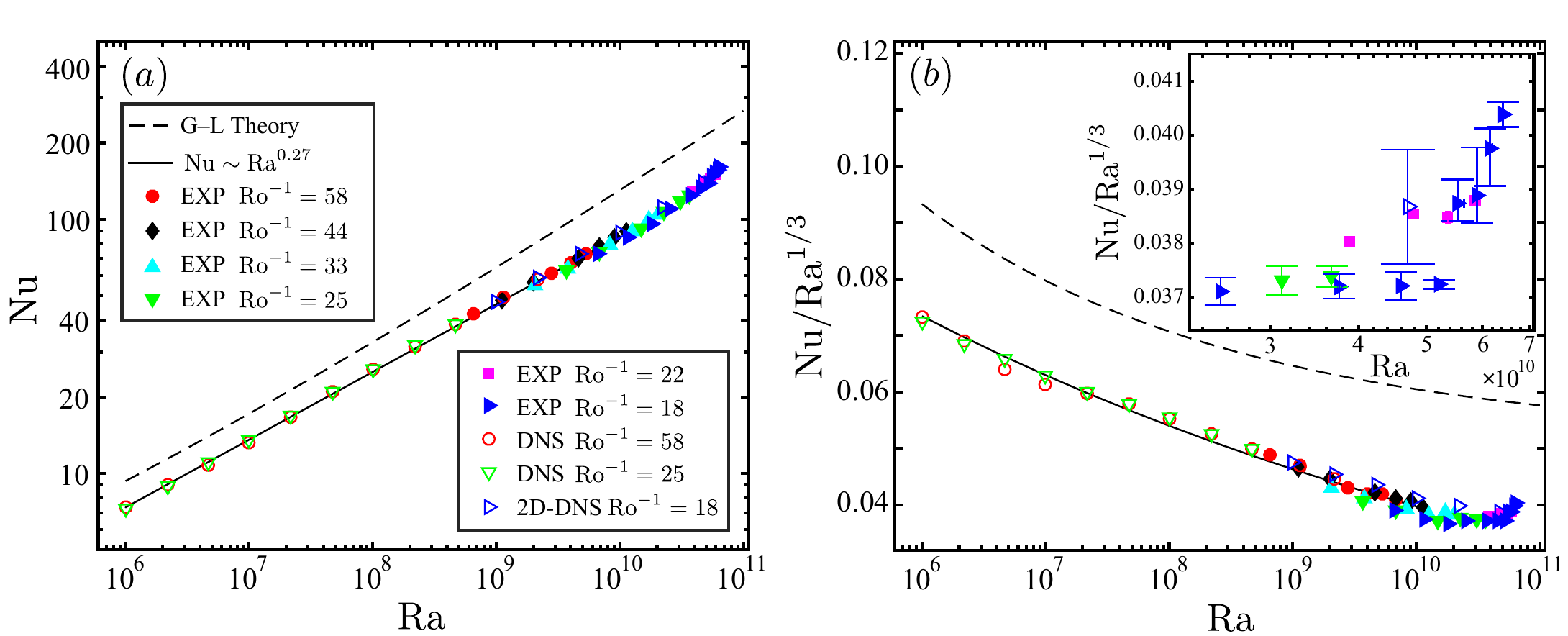}
\caption{{\bf Global heat transport.} (a) Nusselt number ($\text{Nu}$) as a function of $\text{Ra}$ from experiments (the solid symbols), DNS (the open symbols) in ACRBC and the prediction from Grossmann-Lohse (G--L) theory\cite{gro00} in classical RBC (dashed line). (b) The same plots as (a), but the vertical axis is compensated. Inset: an enlarged portion of the compensated plot at the large $\text{Ra}$ regime, which shows the transition of the effective scaling exponent ($\text{Nu} \propto \text{Ta}^{\gamma}$) to $\gamma > 1/3$.}
\label{fig:heat}
\end{figure*}

\subsection{Global turbulent heat transport}
As shown in the section above, $\text{Nu}$ does not depend much on $\text{Ro}^{-1}$ for high $Ro^{-1}$ regime, in which we will study how heat transfer depends on $\text{Ra}$ in this section. We have performed 48 experiments and more than 130 numerical simulations. Fig.~\ref{fig:setup}(c) shows the parameter space of the experimental and numerical studies. In the experimental studies, we note that the existence of Earth's gravity and lids is unavoidable. Several numerical simulations have been performed to study the influences of Earth's gravity and lids, which show that their effects on $\text{Nu}$ are small. 
 In addition, we note that the centrifugal force increases linearly with radial location. To study the influence of nonuniform driving force, we have performed two sets of simulations with the radial-dependent centrifugal acceleration $\omega^2 r$ and with a constant artificial acceleration $\omega^2 (R_o+R_i)/2$. The results show that the effect of the radial-dependent gravity has a similar role as the non-Oberbeck-Boussinesq effect \citep{gro00,ahl06,sug09}, which does not have much influence on the heat transport and flow structures in the current parameter regime (see the Supplementary Materials for details).

Fig.~\ref{fig:heat} (a) shows the measured $\text{Nu}$ as a function of $\text{Ra}$ for different $\text{Ro}$. Each measurement lasts at least 4 hours after the system has reached a thermally steady situation to ensure a statistically stationary state (for the detailed measurement procedure, see the Supplementary Materials). As being evident in Fig.~\ref{fig:heat}(a), the experiments and numerical simulations are in excellent agreement. Combining experiments and simulations, this study covers more than four decades of $\text{Ra}$, \textit{i.e.} from $10^6$ to $6.5\times 10^{10}$. The range of $\text{Ro}^{-1}$ is from 18 to 58 in experiments, and the data series show a consistent dependence of $\text{Nu}$ on $\text{Ra}$. For comparison, we also plot the data calculated from Grossmann--Lohse (G--L) theory \cite{gro00} in the corresponding $\text{Ra}$ range. 
To better reveal the local exponent, we plot the compensated data in Fig.~\ref{fig:heat}(b). The experimental and numerical data have a lower amplitude as compared to the classical RBC (G--L line) due to the different flow geometry. However, the scaling dependence of $\text{Nu}$ versus $\text{Ra}$ shows a good agreement between the data and G-L theory at $\text{Ra} \lessapprox 10^{10}$ with a scaling exponent $\gamma=0.27\pm0.01$, which is close to the typical value found in two-dimensional RBC. 
\begin{figure*}[ht]
\centering
\includegraphics[width=\linewidth]{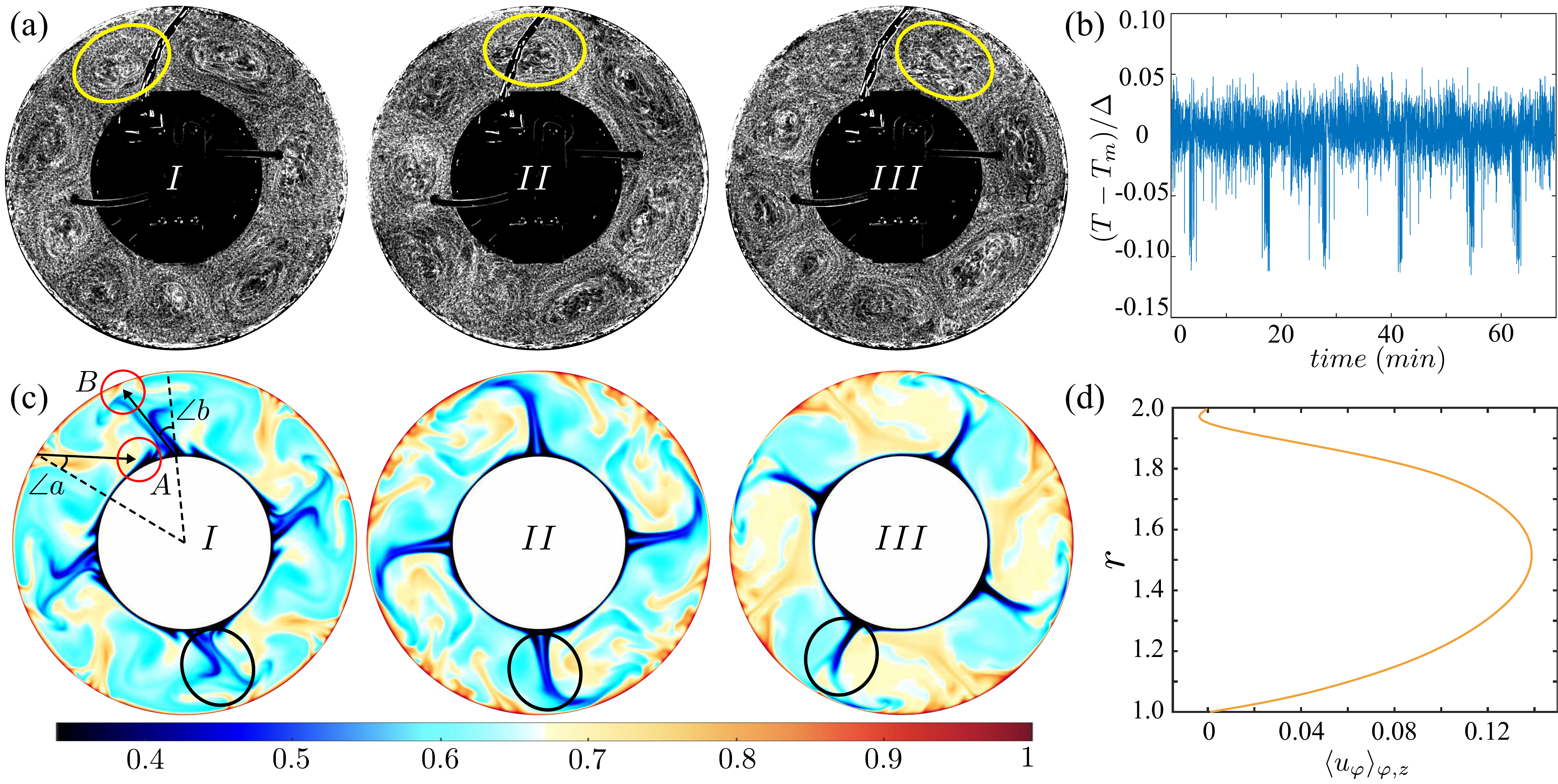}
\caption{{\bf Revolution of convective rolls.} (a, b) Experiments at $\text{Ra}=1.8\times 10^9$ and $\text{Ro}^{-1}=13.8$. (a) Snapshots of streak images revealing the flow patterns. Seeing from the top, the whole system rotates clockwise in experiments. The corresponding movie is available in movie S2. (b) Time series of local temperature fluctuations. The measured point locates \unit{30}{\milli\meter} away from the cold inner cylinder and at mid-height. (c, d) Simulations at $\text{Ra}=10^7$, and $\text{Ro}^{-1}=1$: (c) Snapshots (Top view) of instantaneous temperature fields. (d) The averaged azimuthal velocity profile along the radial direction. The reference of the frame is on the clockwise direction, and the corresponding movie is available in the Supplementary Movie.}
\label{fig:flowpattern}
\end{figure*}

It is very unexpected that the local effective exponent $\gamma$ of $\text{Nu} \propto \text{Ra}^{\gamma}$ even exceeds one-third at $\text{Ra} \gtrsim 4\times 10^{10} $ in experiments. 
Is this scaling regime connected to the appearance of ultimate regime of Rayleigh--B\'enard turbulence \cite{kra62,gro11,he12a}? If it is not in the ultimate regime, what causes this steep scaling exponent?  
As our current $\text{Ra}$ is much lower than the transition $\text{Ra}$ for ultimate turbulence observed in the classical RBC, we cannot provide a concrete answer on this question. The possible answers are that (i) because of the different way of the driving and the geometry, the transition Ra to the ultimate regime at the current system is lower than that in the classical RBC. (ii) Another possibility is that the enhanced local slope is due to the emergence of new flow states. As demonstrated in Fig.~\ref{fig:setup} (c), with $\text{Ra}$ increasing, the location of ($\text{Ra},\, \text{Ro}^{-1}$) where the enhanced local effective scaling appears seems to move from regime III to regime II. Thus the flow may change from a two-dimensional flow state to a three-dimensional flow state due to the strong thermal driving, resulting in a higher $\text{Nu}$ in this $\text{Ra}$ range. 
The two-dimensional simulations for high $\text{Ra}$ coincide well with the experiments, indicating that within the $\text{Ra}$ range explored in two-dimensional simulations the flow is still quasi-two-dimensional. Unfortunately, we cannot push the simulations to the Ra regime with the effective exponent larger than one-third, in which the flow may not be in the quasi-two-dimensional state. We incline to anticipate that the changing from a two-dimensional flow state to a three-dimensional flow state may be responsible for the enhancement of local effective exponent at $Ra \gtrsim 4\times10^{10}$. 
We notice that the scaling range with $\gamma > 1/3$ is very narrow in the current work, therefore further work at even higher Ra is needed to clarify this issue.

\subsection{Zonal flow}

We now study the dynamics of zonal flow experimentally and numerically. The "zonal flow" phenomenon has been investigated in experiments of geophysical and astrophysical flows \cite{hei05,har15}. Fig.~\ref{fig:flowpattern}(b) shows the time series of local temperature fluctuations in water at $\text{Ra}=6.6\times 10^9$, $\text{Pr}=4.3$, and $\text{Ro}^{-1}=18$, which unexpectedly shows a noticeable periodicity. To prove that this novel phenomenon is connected to the azimuthal movement of the coherent flow, we perform flow visualization with aqueous glycerol solution as mentioned in Methods to demonstrate the flow pattern.

Fig.~\ref{fig:flowpattern}(a) shows some typical streak images from flow visualization at three different time (for video, see movie S2), where there are four pairs of rolls parallel to the rotating axis. For reference, we highlight one of the convective rolls using a yellow ellipse. From Fig.~\ref{fig:flowpattern} (aI--aIII), unexpectedly the convection rolls move clockwise around the center with a faster rotation rate than the background rotation of the experimental system, which we name "zonal flow". The flow visualization further justifies that the revolution of the cold plume arms triggers the periodic temperature signals measured by the thermistors.

Next to the experiments, we also find this phenomenon in the simulations (see Fig.~\ref{fig:flowpattern}(c) and movie S3).
We use a black ellipse to highlight a selected cold plume. As shown in Fig.~\ref{fig:flowpattern}(c),
on the reference of the (clockwise) rotating frame, the plume arms still evolve in a clockwise direction, indicating a net rotation of the coherent flow (zonal flow).
This zonal flow is further quantified through the averaged (axially and azimuthally in space and over time) azimuthal velocity profile versus the radial distance from the inner cylinder wall $r$, as shown in Fig.~\ref{fig:flowpattern}(d).

What causes this net rotation of the convection rolls? 
The dashed lines in Fig. \ref{fig:flowpattern}(c\uppercase\expandafter{\romannumeral1}) mark the direction of hot/cold plumes without Coriolis force. But because of the effects of Coriolis force, the plumes deflect to their left when the system rotates clockwise. As is evident in the Fig. \ref{fig:flowpattern}(c\uppercase\expandafter{\romannumeral1}) where the plumes are just formed in the beginning stage, the deflected angle of hot plumes ($\angle a$) is approximately equal to the deflected angle of cold plumes ($\angle b$). However, because of the different curvature of the cylinders, the similar deflected angles of the hot and cold plumes induce the different effects. The hot plumes directly affect the position (A) where the cold plumes are ejected, resulting in the clockwise rotation of the cold plumes and flow, whereas the impact of the cold plumes does not directly affect the motion of the hot plumes due to a relatively large distance the impact position of the cold plumes (B) and the ejecting position of the hot plumes. Thus, the hot plumes win and push the overall flow to move in the same direction of the background rotation. 
We also perform the experiments and simulations with the opposite direction of the background rotation, and the results are consistent. In addition, several numerical simulations have been performed to test the dependence of zonal flow on the radius ratio $\eta$, which shows that the zonal flows become weaker and weaker with $\eta$ increasing from 0.4 to 0.9. The difference in the curvature of the two cylinders plays the key role for the net rotation of convection rolls. One may expect to observe different types of zonal flows at different $\eta$, such as two large scale winds with the opposite directions near the plates \cite{har15}. This remarkable net rotation of the coherent flow structures along the same direction of the background rotation could be connected to many relevant flow phenomena in nature, which deserves systematical studies in the future.

\section{Summary}
In summary, we have experimentally and numerically studied the global heat transport and turbulent flow structures in a rotating annulus with hot outer cylinder and cold inner cylinder, \textit{i.e.}, ACRBC. In experiments, the mean centrifugal acceleration covers the range from 5 to 60 times gravitational acceleration. We show that the effective scaling exponent of global heat transport transitions to $\gamma=1/3$ at $\text{Ra} \approx 10^{10}$ and finally exceeds $\gamma > 1/3$ for $\text{Ra} \gtrapprox 4\times 10^{10}$. 
Unexpectedly, we observed the faster azimuthal motion of the coherent overall flow structure faster than the background rotation of the system, which we call zonal flow, and provided a physical understanding on this remarkable net rotation motion of the coherent structures. This novel experimental approach sheds new lights on how to efficiently extend the parameter regimes for the study of buoyancy driven turbulent flows. Furthermore, the findings in the current system driven by the centrifugal acceleration can help us understand phenomena in geophysical and astrophysical flows.

\section{Methods}
{\it Measurement techniques.} In our experiments, all the temperature measurements are based on negative temperature coefficient thermistors. We use $6\frac{1}{2}$-digit multimeter (Keithley, \texttt{2701}) to measure the resistances of the thermistors, and the resistance can be converted to temperature using the Steinhart-Hart equation\cite{lav97}. Two types of thermistors are used: One with a head diameter of \unit{2.5}{\milli\meter} (Omega, \texttt{44131}) is used to measure the temperature of the inner and outer cylinders, and the other with a head diameter of \unit{300}{\micro\meter} (Measurement Specialties, \texttt{GAG22K7MCD419}) is inserted into the convection cell to resolve fast temperature fluctuations, which is located \unit{30}{\milli\meter} away from the cold cylinder and at mid-height in the vertical direction. 

To visualize the flow field, we use nylon fibers with length $l = \unit{3}{\milli\meter}$, and diameter $d =  \unit{0.5}{\milli\meter}$. Nylon has a density of $\rho_{p}\approx \unit{1.14}{\gram\per\centi\meter\cubed}$ that is a little heavier than water. In the flow visualization experiments, we use aqueous glycerol solution with 54\% glycerol by volume to match the density of nylon. The properties of the aqueous glycerol solution are listed in the Supplementary Materials for reference. Four light-emitting diods are used as light source, and a charge-coupled device camera (Ximea, \texttt{MD028MU-SY}) is put on the top of the cell to record images. As the system rotates rapidly, it is difficult to make the camera rotate synchronously. So we keep the camera fixed, and then set the frame rate of the camera the same as the rotational speed. Processing the images taken by camera through a MATLAB code, we can get streak images to visualize the flow field. More processing details can be found in the Supplementary Materials.

\bigskip


\begin{acknowledgments}
\textbf{Acknowledgments:} We thank Y. Yang, Y. Liu, Q. Zhou, and S. Liu for insightful discussions, and thank G.-W. Bruggert for the technical assistance with the setup. 

\bigskip

\textbf{Funding:} This work was financially supported by the Natural Science Foundation of China under grant nos. 11988102, 91852202, 11861131005, and 11672156; the Tsinghua University Initiative Scientific Research Program (20193080058); and the Netherlands Organisation for Scientific Research (NWO) through the Multiscale Catalytic Energy Conversion (MCEC) research center. X.Z. acknowledges the financial support from the Center for Mathematical Sciences and Applications at the Harvard University.

\bigskip

\textbf{Author contributions:}
C.S. conceived the project. H.J. and D.W. conducted the experiments. X.Z. and H.J. performed the simulations. H.J., S.G.H., and C.S. designed the setup. All authors discussed the physics and contributed to the writing of the manuscript.

\bigskip

\textbf{Competing interests}: The authors declare that they have no competing interests.

\bigskip

\textbf{Data and materials availability:}: All data needed to evaluate the conclusions in the paper are present in
the paper and/or the Supplementary Materials. Additional data related to this paper may
be requested from the authors.

\end{acknowledgments}
\end{document}